\documentclass[%
 reprint,
 amsmath,amssymb,
 aps,
 prl,
 floatfix,
]{revtex4-2}
\usepackage{lineno}
\usepackage{graphicx}
\usepackage{dcolumn}
\usepackage{bm}
\usepackage[colorlinks=true, linkcolor=blue, citecolor=blue, urlcolor=blue]{hyperref}
\usepackage{booktabs}   
\usepackage{siunitx}    
\usepackage{braket} 

\begin{document}

\title{Observation of Curvature-Controlled Spatial Phase Separation \\
in a Momentum-State Lattice}

\author{Chen Chen}
\affiliation{Advanced Materials, Hong Kong University of Science and Technology (Guangzhou), Guangzhou 511453, China}
\author{Peng Zhu}
\affiliation{Advanced Materials, Hong Kong University of Science and Technology (Guangzhou), Guangzhou 511453, China}
\author{Shuchang Ma}
\affiliation{Advanced Materials, Hong Kong University of Science and Technology (Guangzhou), Guangzhou 511453, China}
\author{Dezhou Deng}
\affiliation{Advanced Materials, Hong Kong University of Science and Technology (Guangzhou), Guangzhou 511453, China}
\author{Peng Chen}
\email{pengchen@hkust-gz.edu.cn}
\affiliation{Advanced Materials, Hong Kong University of Science and Technology (Guangzhou), Guangzhou 511453, China}
\affiliation{Quantum Science Center of Guangdong-Hong Kong-Macau Greater Bay Area, Shenzhen 518045, China}

\date{\today}

\begin{abstract}
We demonstrate spatial control of quasiperiodic transport in a momentum-state lattice of $^{87}$Rb atoms. Programmable power-law hopping realizes the curved-spacetime Aubry--Andr\'e--Harper model and suppresses wave-packet spreading at progressively weaker quasiperiodic modulation. Position-resolved injection reveals a progression from subextended spreading to localized confinement within the same target Hamiltonian. At intermediate positions, varying the quasiperiodic phase switches the wave packet between confinement and spreading across several sites, providing dynamical evidence consistent with the predicted swing regime. Changing the hopping exponent displaces this phase-sensitive crossover, in semiquantitative agreement with a finite-time criterion fixed at a single exponent. These results establish the spatial profile of kinetic energy as a programmable control parameter for organizing transport and confinement in quasiperiodic systems.
\end{abstract}

\maketitle

Anderson localization---the suppression of quantum transport by disorder---is among the most studied phenomena in physics~\cite{Anderson1958,Evers2008,Abrahams1979,Lee1985}. In one-dimensional quasiperiodic systems, the Aubry-Andr\'e-Harper (AAH) model provides a paradigmatic realization: the localization transition occurs simultaneously for all eigenstates at the self-dual point $V_0/J_0=2$, and no mobility edge exists~\cite{Harper1955,Aubry1980,Kohmoto1987,Hiramoto1989}.  Breaking self-duality through modified potentials~\cite{Ganeshan2015,Biddle2010,Wang2020Mosaic,Li2017}, long-range couplings~\cite{DasSarma1988,Deng2019PowerLaw}, or non-Hermitian terms~\cite{Liu2020NonHermitian}---can introduce energy-dependent mobility edges~\cite{Luschen2018,An2021GAA,Wang2022AAH,Semeghini2015,Lahini2009}, most experimental realizations rely on a spatially uniform kinetic energy scale~\cite{An2021GAA,Wang2022AAH}. With spatially uniform hopping, the ratio $V_0/J_0$ is position independent. Spatially varying hopping introduces an additional control by making the kinetic-energy scale depend on position. This raises the question of how transport and confinement become organized within a single lattice and how their spatial crossover can be controlled.

Lattice models of curved spacetime provide a framework for engineering such kinetic-energy landscapes through spatially varying couplings~\cite{Boada2011,Morice2021Horizons,Kedem2020BlackWhite}. Related synthetic geometries have been explored in analogue-gravity platforms~\cite{Unruh1981,Hawking1974,Barcelo2011,Steinhauer2016,Drori2019,Lahav2010,Weinfurtner2011,Philbin2008}, expanding-universe dynamics~\cite{Viermann2022}, and curved-space wave propagation~\cite{Morice2022Dynamics,Sheng2021Escape,He2024,Wang2020NSR,Mertens2022Thermalization,Konye2022Horizons,RodriguezLaguna2017}. The curved-spacetime AAH (CST-AAH) model, in which the hopping follows a power law, identifies a local self-dual onset and predicts a position-space phase separation in which subextended, swing, and localized regions coexist~\cite{Li2023CSTAAH}. The phase boundary depends on the quasiperiodic phase $\phi$, giving rise to a swing mobility edge whose spectral signatures have been characterized numerically~\cite{Li2023CSTAAH,Mao2025CSTAAH}. These predictions motivate experimental measurements that resolve both the position and phase dependence of transport.

In this work, we realize the CST-AAH model in a 21-site momentum-state lattice of $^{87}$Rb atoms~\cite{Gadway2015,Meier2016PRA,Meier2018,Yuan2023Momentum,An2018}, using individually calibrated Bragg couplings to program the power-law hopping profile $J_n\propto(n+1)^{-\sigma}$. Mapping the $(\sigma,V_0/J_0)$ plane shows that curvature strongly enhances transport suppression, allowing weaker quasiperiodic modulation to inhibit wave-packet spreading. Site-resolved injection along the hopping gradient reveals a position-dependent crossover from transport to confinement at fixed $\sigma$ and $V_0/J_0$, with a phase-sensitive crossover consistent with the predicted swing regime. These results extend quasiperiodic localization from energy-resolved mobility edges in uniform lattices to position-resolved crossovers in inhomogeneous lattices, establishing spatial geometry as a programmable control parameter.

\begin{figure}[htbp]
    \centering
    \includegraphics[width=\linewidth]{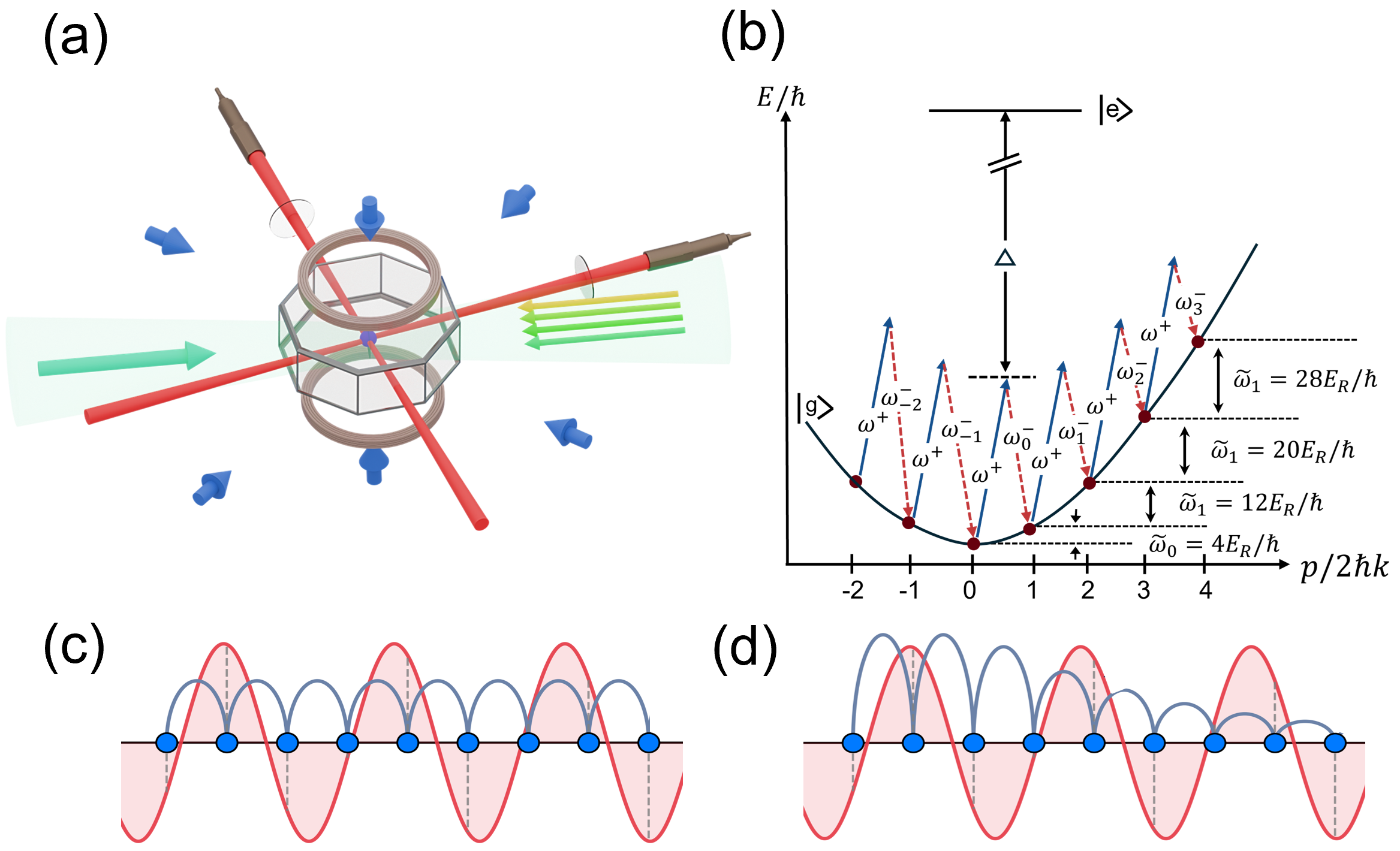}
    \caption{
        Experimental platform and CST-AAH model. (a) A $^{87}$Rb Bose-Einstein condensate is driven by counter-propagating Bragg beams. (b) Momentum states $|n\rangle=|2n\hbar k\rangle$ form a synthetic lattice with programmable nearest-neighbor couplings. (c) Flat AAH model with uniform hopping. (d) CST-AAH model with power-law hopping decay and quasiperiodic on-site potential.
    }
    \label{fig1}
\end{figure}

The Hamiltonian combines the quasiperiodic AAH model with a spatially varying hopping profile that encodes a synthetic curved geometry, given by
\begin{equation}
\hat{H} = \sum_{n=0}^{L-2} J_n \left( \hat{c}_{n+1}^\dagger \hat{c}_n + \text{H.c.} \right)
- \sum_{n=0}^{L-1} V_n \hat{c}_n^\dagger \hat{c}_n ,
\label{eq:hamiltonian}
\end{equation}
where $\hat{c}_n^\dagger$ creates an atom on synthetic lattice site $n$ and $L$ is the number of sites. The quasiperiodic potential is $V_n=V_0\cos(2\pi\beta n+\phi)$ with $\beta=(\sqrt{5}-1)/2$ and $\phi=0$ by default.  Curvature is encoded in the nearest-neighbor hopping
\begin{equation}
J_n=J_0\left(\frac{1}{n+1}\right)^\sigma ,
\label{J_n}
\end{equation}
where $\sigma$ indicates warping degree of spacetime. For $\sigma=0$, the model reduces to the flat AAH chain with uniform hopping [Fig.~\ref{fig1}(c)].  For $\sigma>0$, $J_n$ decreases with $n$ [Fig.~\ref{fig1}(d)], so that the local competition between tunneling and quasiperiodic modulation varies monotonically along the chain. 

Figure 1(a) illustrates our experimental platform. The atoms are confined in a crossed optical dipole trap~(XODT) formed by two 1064~nm laser beams with waist sizes of 55 and 70~$\mu$m, respectively. We first prepare a $^{87}$Rb Bose-Einstein condensate~(BEC) containing $4 \times 10^5$ atoms at a temperature of $\sim$100~nK, with trap frequencies of $\omega/2\pi = (62, 111, 102)$~Hz. The momentum lattice is implemented by a pair of counter-propagating 1064~nm lattice beams, which drive two-photon Bragg transitions between momentum states $|n\rangle = |2n\hbar k\rangle$ [Fig.~\ref{fig1}(b)]. The lattice beams have a waist of $350~\mu$m at the BEC position, and their axis is oriented at $9^\circ$ relative to one of the dipole-trap beams. The frequency and amplitude of the counter-propagating beam are precisely controlled via two acousto-optic modulators~(AOMs). To minimize momentum spread and suppress breathing and sloshing  modes during the transfer, a 0.4~s ramp of decreasing XODT depth and increasing lattice power is applied, followed by a 0.1~s hold. This preparation sequence yields a pure BEC of $2 \times 10^5$ atoms at $\sim$10~nK, while elongating the cloud along the lattice axis with reduced mean-field interactions, providing a well-controlled initial state for the momentum lattice experiments.

We implement a synthetic lattice of 21 momentum sites coupled by 20 nearest-neighbor links. The incident beam contains a single optical frequency, while the counter-propagating beam carries a multi-frequency waveform generated by an arbitrary waveform generator. Each frequency component addresses a specific nearest-neighbor transition, where its amplitude and detuning dictate the hopping strength $J_n$ and the onsite potential difference $\Delta V_n$ between adjacent sites, respectively. This frequency-resolved control enables independent engineering of the nearest-neighbor hopping amplitudes and onsite potentials, realizing the full Hamiltonian in Eq.~(\ref{eq:hamiltonian}). 

For each data set, the voltage-to-hopping response is calibrated under the corresponding multi-tone background. In standard runs, the multi-frequency lattice waveform is applied for 1.5~ms. After the lattice driving, all dipole trap and lattice beams are switched off. Momentum components are resolved by a 15~ms time-of-flight expansion, and the site populations $P_n$ are extracted from subsequent absorption imaging.

\begin{figure}[htbp]
    \centering
    \includegraphics[width=\linewidth]{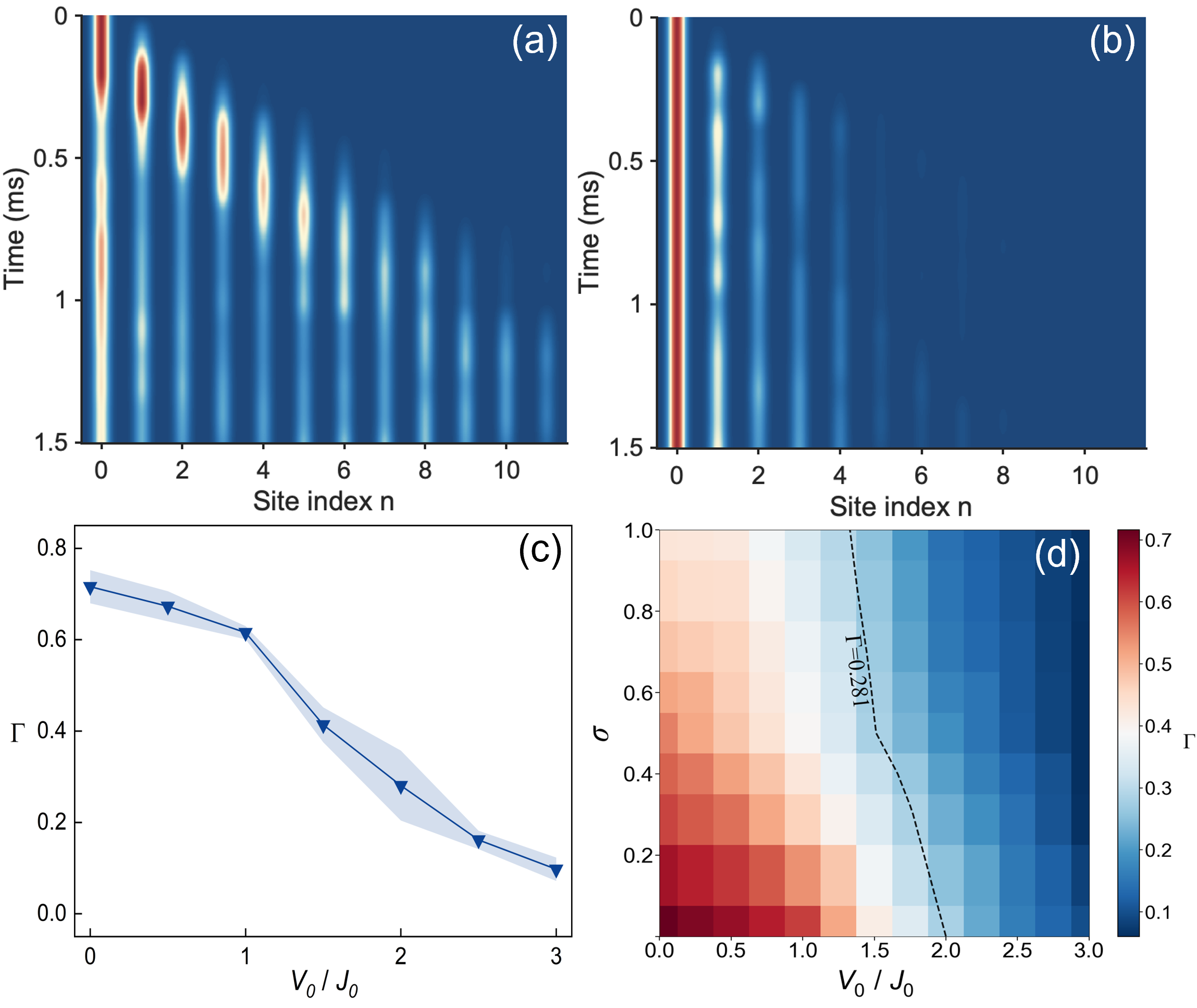}
    \caption{
    Flat-lattice benchmark and curvature-induced localization. (a,b)~Measured site populations versus evolution time for the flat AAH lattice at $V_0/J_0=0$ and $V_0/J_0=2$. (c)~Fractal dimension $\Gamma$ extracted at $t=1.5$~ms versus $V_0/J_0$ for $\sigma=0$. Shaded band: $\pm1$ standard deviation over 10 runs. (d)~Experimentally measured $\Gamma$ over the parameter space $(V_0/J_0,\sigma)$, averaged over 10 independent runs. The dashed contour marks the iso-fractal-dimension condition $\Gamma=\tilde{\Gamma}=0.281$, fixed by the flat-lattice self-dual benchmark.}
    \label{fig2}
\end{figure}

Figures~\ref{fig2}(a,b) show the measured time evolution of the site populations for an initial state localized at $n=0$ in the flat AAH lattice, with $J_0/h=1000~\mathrm{Hz}$. At $V_0/J_0=0$, the cloud expands across the available lattice  within the observation window. At $V_0/J_0=2$, it remains concentrated near the origin after the same evolution period. These results validate the reliability of the frequency-resolved addressing, amplitude calibration, and population extraction protocols before introducing the hopping gradient.

Following Refs.~\cite{Hiramoto1992,Li2023CSTAAH}, we characterize the spatial extent of the measured wave packet using the fractal dimension, defined as $\Gamma=-{\ln(\mathrm{IPR})}/{\ln L}$, with $\mathrm{IPR}=\sum_n P_n^2$, where $0\leq\Gamma\leq1$ and a larger $\Gamma$ corresponds to a more extended distribution across the lattice. As shown in Fig.~\ref{fig2}(c), $\Gamma$ decreases monotonically with $V_0/J_0$~\cite{An2021GAA}. We therefore use $\tilde{\Gamma}\equiv\Gamma(\sigma=0,V_0/J_0=2)=0.281$, extracted at the flat-lattice self-dual point, as a reference value for the comparisons below.

With the flat-lattice response established, we now examine how a spatial hopping gradient modifies transport. In the flat AAH lattice, uniform hopping makes $V_0/J_0$ a position-independent ratio between the quasiperiodic modulation amplitude and the kinetic-energy scale. For $\sigma>0$, the hopping decreases toward larger site indices, so that the local ratio $V_0/J_n$ increases along the chain. As the wave packet spreads, it therefore encounters regions in which quasiperiodic modulation becomes increasingly important relative to tunneling. This spatial variation can suppress further propagation even when the global ratio $V_0/J_0$ remains below the flat-lattice self-dual value.

To examine this expectation experimentally, we map $\Gamma$ over the $(\sigma,V_0/J_0)$ plane at $t=1.5~\mathrm{ms}$ [Fig.~\ref{fig2}(d)]. As $\sigma$ increases, the low-$\Gamma$ region extends systematically toward smaller $V_0/J_0$, with the reference contour $\Gamma=\widetilde{\Gamma}$ tracing the displacement of the transport crossover. At $\sigma=1$, strong suppression of spreading already occurs around $V_0/J_0\simeq1$, approximately half the flat-lattice self-dual value. These measurements show that the spatial hopping profile controls the modulation strength required to inhibit transport over the observation time. We next resolve how this response is organized along the lattice by varying the injection position and quasiperiodic phase.
We now focus on $\sigma=1.0$ and $V_0/J_0=1.0$, where the CST-AAH model is expected to support the coexistence of subextended, swing, and localized regimes. These regimes are distinguished most directly by their dependence on the quasiperiodic phase: localized and subextended states retain their respective localization characters as $\phi$ is varied, whereas states in the swing regime switch between localized-like and subextended-like behavior. This phase sensitivity provides a particularly discriminating experimental signature of the swing regime, because it can distinguish phase-dependent switching from a phase-insensitive suppression of transport caused simply by a reduced local hopping scale.

\begin{figure*}[htbp]
    \centering
    \includegraphics[width=0.9\linewidth]{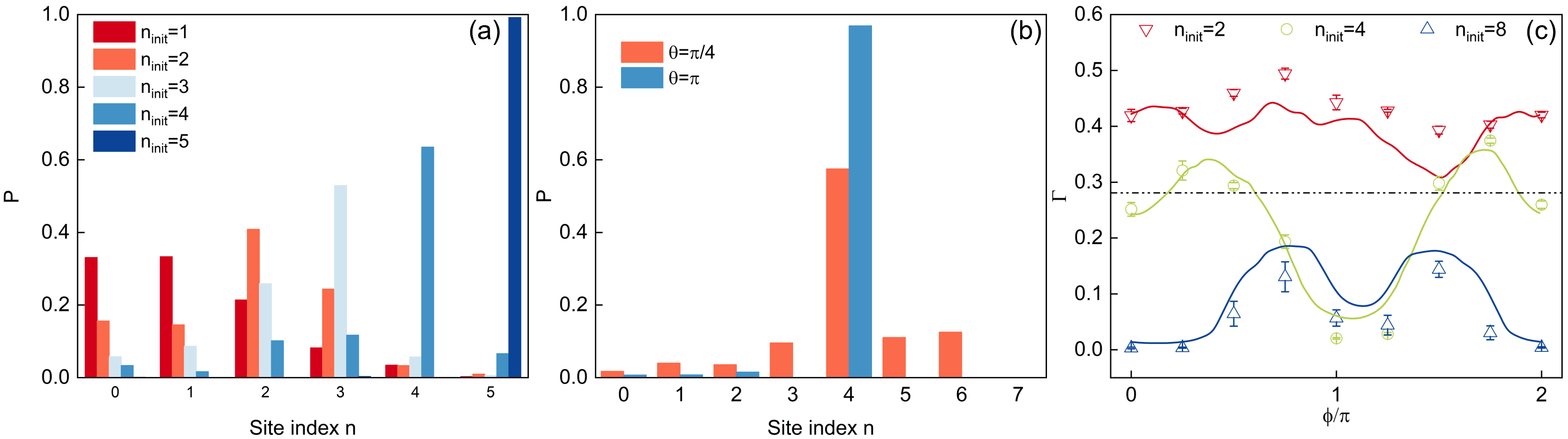}
    \caption{
    Observation of position-space phase separation at $\sigma=1.0$ and $V_0/J_0=1.0$.
    (a) Population distributions $P(n)$ for injection sites $n_{\mathrm{init}}=1$--$5$ at $\phi=0$.
    (b) Population distributions for $n_{\mathrm{init}}=4$ at $\phi=\pi/4$ and $\pi$. The phase-dependent redistribution extends beyond the injection site and its nearest neighbors, with appreciable population at $n=6$ for $\phi=\pi/4$.
    (c) Phase dependence of $\Gamma$ for $n_{\mathrm{init}}=2$, $4$, and $8$ over $\phi=0$--$2\pi$. Open symbols denote the experimental data with $\pm1$ standard-deviation error bars, and solid curves show the numerical simulations including mean-field interactions and pure dephasing. The dash-dotted horizontal line marks $\widetilde{\Gamma}=0.281$. }
    \label{fig3}
\end{figure*}

To establish the spatial structure, we first perform a position scan at fixed quasiperiodic phase. We keep the condensate in the same physical momentum state and synchronously translate the programmed hopping and quasiperiodic onsite profiles. In the translated coordinates, this maps the condensate onto different effective injection sites $n_{\mathrm{init}}$ of the same target CST-AAH Hamiltonian while preserving $\phi$. Figure~\ref{fig3}(a) shows a systematic crossover from spreading to confinement. At $n_{\mathrm{init}}=1$ and $2$, atoms spread over multiple sites, whereas the distributions progressively narrow at $n_{\mathrm{init}}=3$ and $4$, becoming strongly confined at $n_{\mathrm{init}}=5$.

At the intermediate injection site $n_{\mathrm{init}}=4$, we compare the population distributions at $\phi=\pi/4$ and $\pi$ [Fig.~\ref{fig3}(b)]. For $\phi=\pi/4$, the wave packet spreads across several sites, with appreciable population reaching $n=6$, whereas for $\phi=\pi$ it remains concentrated at the injection site. The total population outside the three-site region $n=3,4,5$ decreases from $(21.9\pm1.3)\%$ to $(3.12\pm0.21)\%$. The phase-controlled switching thus involves multisite transport beyond the injection site and its nearest neighbors.

We next scan the quasiperiodic phase $\phi$ over $0$--$2\pi$ at three representative injection sites [Fig.~\ref{fig3}(c)]. Across the sampled phases, $\Gamma$ remains above $\tilde{\Gamma}$ at $n_{\mathrm{init}}=2$ and well below it at $n_{\mathrm{init}}=8$, indicating subextended-like and localized-like responses, respectively. At $n_{\mathrm{init}}=4$, $\Gamma$ crosses $\tilde{\Gamma}$ as $\phi$ is varied. In the single-particle picture, a localized injection samples multiple eigenstates, linking the finite-time spreading to their spatial structure and interference. Together with the multisite population redistribution, this position-dependent phase response provides dynamical evidence consistent with the predicted swing regime~\cite{Li2023CSTAAH}. The coherent single-particle calculation, however, overestimates the modulation of $\Gamma(\phi)$ at $n_{\mathrm{init}}=2$ and predicts a $\pi$ periodicity that is broken in the measured phase response.

Previous studies of momentum-state lattices have shown that transport is sensitive to both the loss of coherence between momentum orders~\cite{An2017,Paladugu2024} and interaction-induced shifts of the site energies~\cite{An2018PRX,Li2023}. Motivated by these findings, we incorporate effective dephasing and a population-dependent mean-field shift into the dynamics. The density matrix $\rho$ then obeys
\begin{equation}
  \frac{d\rho}{dt}
  = -\frac{i}{\hbar}[H_{\rm eff}[\rho],\rho]
    + \sum_n \left(
      L_n \rho L_n^\dagger
      - \tfrac{1}{2}\{L_n^\dagger L_n,\rho\}
    \right),
  \label{eq:master}
\end{equation}
with
\begin{equation*}
  H_{\rm eff}[\rho]
  = H - U\sum_n \rho_{nn}\ket{n}\!\bra{n},
  \qquad
  L_n = \sqrt{\gamma_\phi}\ket{n}\!\bra{n}.
\end{equation*}
Here $H$ is the single-particle lattice Hamiltonian, $-U\rho_{nn}$ is the mean-field shift associated with the instantaneous site population $\rho_{nn}$~\cite{An2018PRX,Li2023}, and $\gamma_\phi$ is the decay rate of the intersite coherences. The solid curves in Fig.~\ref{fig3}(c) use $U/h=800~\mathrm{Hz}$, estimated from the peak atomic density and the $^{87}$Rb $s$-wave scattering length, and $\gamma_\phi=1000~\mathrm{s^{-1}}$, determined independently from the decay envelope of measured Rabi oscillations.

Dephasing damps the intersite coherences that mediate transport between momentum states. Spatial separation of the momentum components, finite momentum width, inhomogeneous Bragg detunings, and fluctuations of the Bragg couplings can all contribute to this loss of coherence. For $\sigma=1$, the larger hopping amplitudes on the low-index side allow the wave packet to spread over several sites during the evolution. At $n_{\mathrm{init}}=2$, dephasing reduces the contribution of coherent interference to transport and weakens the calculated modulation of $\Gamma(\phi)$, bringing it closer to the measured response.

For single-site injection, pure dephasing preserves the $\pi$ periodicity of the noninteracting population dynamics. Interactions can break this symmetry through the mean-field shift $U\rho_{nn}$, which arises from $s$-wave collisions and evolves with the instantaneous site populations. Under $\phi\to\phi+\pi$, the quasiperiodic potential reverses sign, whereas the sign of the mean-field term remains unchanged. This nonlinear feedback allows the wave packets at $\phi$ and $\phi+\pi$ to evolve into different population distributions, consistent with the observed departure of $\Gamma(\phi)$ from $\pi$ periodicity.

\begin{table}[htbp]
\centering
\caption{Phase-switching score $S(n,\sigma)$ for representative curvatures $\sigma=1$, $0.75$, and $0.6$ at $V_0/J_0=1$. Here, $p_{\rm loc}$ is the fraction of sampled phases satisfying $\Gamma<\widetilde{\Gamma}=0.281$, and $S=4p_{\rm loc}(1-p_{\rm loc})$. The maximum of $S$ identifies the experimental phase-sensitive crossover position $n_{\Gamma}^{(\mathrm{exp})}$, with broad maxima indicating a crossover interval. }
\label{tab1}
\begin{tabular}{
    S[table-format=2.0]  
    S[table-format=1.2]   
    S[table-format=2.0]   
    S[table-format=1.2]
    S[table-format=2.0]   
    S[table-format=1.2]
}
\toprule\toprule
\multicolumn{2}{c}{$\sigma=1$} & 
\multicolumn{2}{c}{$\sigma=0.75$} & 
\multicolumn{2}{c}{$\sigma=0.6$} \\
\cmidrule(r){1-2} \cmidrule(r){3-4} \cmidrule(r){5-6}
{$\text{Site index } n$} & {S} & 
{$\text{Site index } n$} & {S} & 
{$\text{Site index } n$} & {S} \\
\midrule
2 & 0.00 & 4 & 0.75 & 8  & 0.75 \\
3 & 0.75 & 5 & 0.75 & 9  & 0.75 \\
4 & 1.00 & 6 & 1.00 & 10 & 1.00 \\
5 & 0.44 & 7 & 0.75 & 11 & 0.75 \\
6 & 0.44 & 8 & 0.75 & 12 & 0.75 \\
\bottomrule\bottomrule
\end{tabular}
\end{table}

\begin{figure}[htbp]
    \centering
    \includegraphics[width=0.8\linewidth]{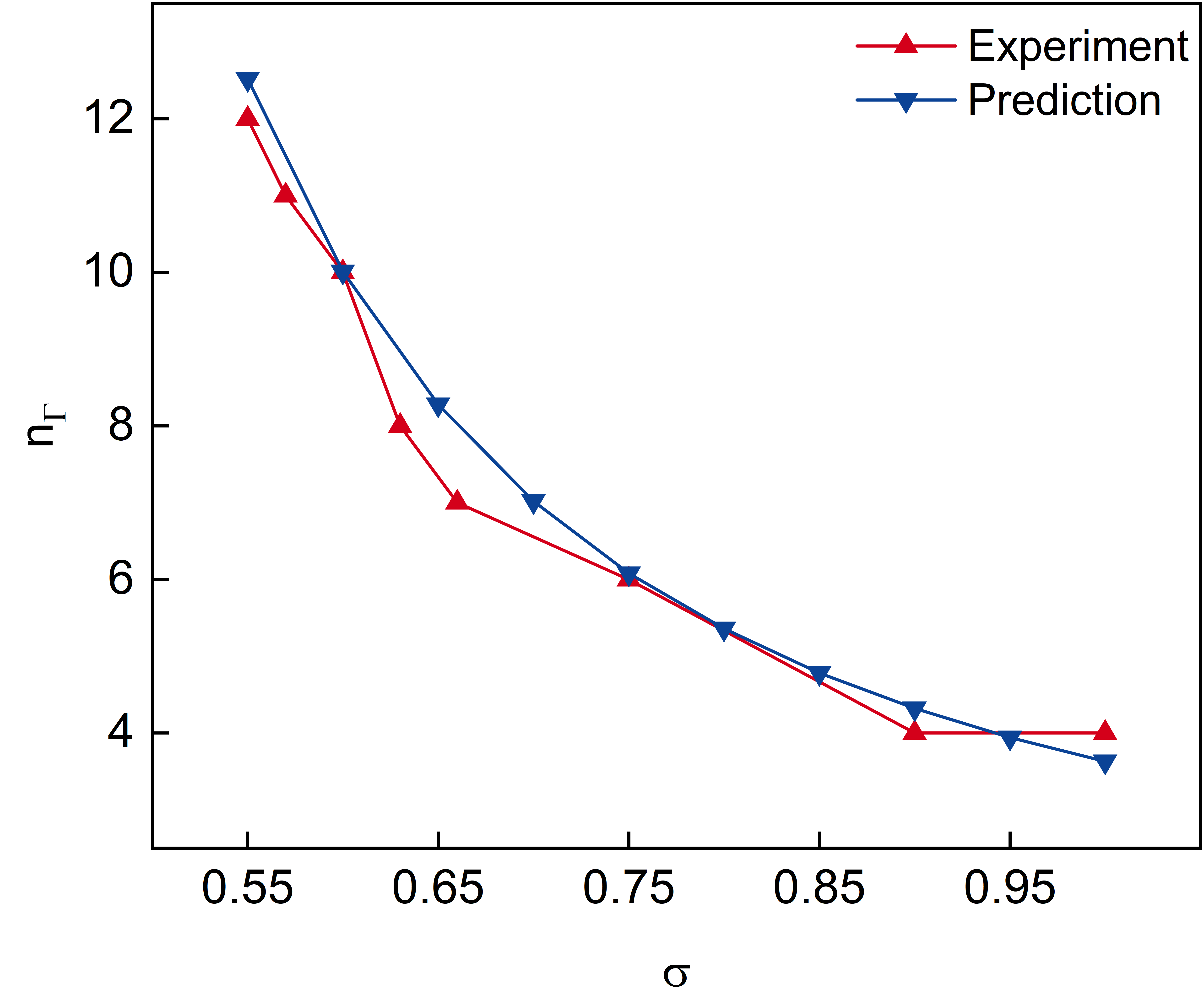}
    \caption{Curvature-controlled displacement of the phase-sensitive crossover. Experimentally extracted crossover positions $n_{\Gamma}^{\mathrm{(exp)}}$ (red upward triangles) and the predictions $n_{\Gamma}^{\mathrm{(pred)}}$ (blue downward triangles) from the calibrated finite-time model are shown as functions of $\sigma$.
}
    \label{fig4}
\end{figure}

Having established the phase-sensitive crossover at $\sigma=1$, we next demonstrate that its spatial position can be tuned by varying the curvature exponent. Decreasing $\sigma$ weakens the spatial dependence of the programmed hopping, such that $J_n$ decreases more slowly with increasing site index and the phase-sensitive crossover shifts toward larger $n$. We probe this shift by repeating the position- and phase-resolved measurements for several smaller values of $\sigma$.

To quantify the phase-sensitive interface without dataset-dependent normalization, we define the fraction of sampled phases exhibiting a localized response, $p_{\rm loc}(n,\sigma)=\sum_{k=1}^{N_\phi}\Theta\left[\widetilde{\Gamma}-\Gamma(n,\sigma,\phi_k)\right]/{N_\phi}$, where $\Theta(x)$ is the Heaviside step function, so that each sampled phase contributes unity when $\Gamma<\widetilde{\Gamma}$ and zero when $\Gamma>\widetilde{\Gamma}$. The corresponding switching score is $S(n,\sigma)=4p_{\rm loc}(n,\sigma)\left[1-p_{\rm loc}(n,\sigma)\right]$, which satisfies $0\leq S\leq1$ and reaches unity when localized and subextended responses occur with equal phase weight. We identify the experimental phase-sensitive crossover position as
$n_{\rm \Gamma}^{(\rm exp)}(\sigma)=\operatorname*{arg\,max}_{n}S(n,\sigma)$. For a broad maximum, the corresponding neighboring sites are retained as the crossover interval.

As shown in Table~\ref{tab1}, the crossover position moves systematically toward larger $n$ as $\sigma$ is reduced, from $n\simeq4$ at $\sigma=1$ to $n\simeq6$ at $\sigma=0.75$ and $n\simeq10$ at $\sigma=0.6$. For a quantitative site-resolved prediction, we instead use the site-centered ratio $\Lambda_{\rm site}(n,\sigma)=V_0/(J_{n-1}+J_n)$, which accounts for the two bonds adjacent to the injection site. Using the calibrated finite-time model of Fig.~\ref{fig3}, we determine the phase-switching ratio from $p_{\rm loc}^{(\rm sim)}(\Lambda_{\rm ps})={1}/{2}$, yielding $\Lambda_{\rm ps}\simeq2.07$ (see Supplemental Material).

The crossover position is then fixed by $\Lambda_{\rm site}\left(n_{\rm \Gamma}^{(\rm pred)},\sigma\right)=\Lambda_{\rm ps}.$ For Eq.~\ref{J_n}, the smooth-gradient approximation
$J_{n-1}+J_n\simeq2\sqrt{J_{n-1}J_n}$ gives, for $V_0/J_0=1$,
\begin{equation}
n_{\Gamma}^{(\rm pred)}(\sigma)=\frac{\sqrt{1+4(2\Lambda_{\rm ps})^{2/\sigma}}-1}{2}.
\end{equation}

Figure~\ref{fig4} compares the measured and predicted crossover positions. Fixing $\Lambda_{\rm ps}$ from the simulated phase response at $\sigma=1$, we predict the crossover displacement at smaller $\sigma$ without further adjustment. The semiquantitative agreement links the measured position of the phase-sensitive crossover to the programmed hopping profile.



The curvature-controlled displacement demonstrated above establishes a position-resolved dynamical crossover that complements the energy-resolved mobility edges of generalized AAH models~\cite{Ganeshan2015,An2021GAA,Luschen2018}, in which extended and localized eigenstates coexist in different spectral regions. In the CST-AAH lattice, the hopping gradient instead makes the competition between tunneling and quasiperiodic modulation coordinate dependent, so that wave packets injected at different effective sites exhibit distinct transport responses under the same Hamiltonian. The resulting spatial organization may therefore be viewed as a dynamical analogue of a metal--insulator junction realized within a single Hamiltonian~\cite{Song2019Junction}, with the crossover generated continuously by the programmed kinetic-energy landscape.

In conclusion, we have realized the CST-AAH model in a programmable momentum-state lattice. Mapping the fractal dimension shows that increasing hopping inhomogeneity shifts the transport-suppression crossover toward weaker quasiperiodic modulation, while site- and phase-resolved injection reveals a spatial progression from subextended spreading, through a swing phase intermediate response, to localized confinement. Varying the curvature further shifts this phase-sensitive crossover along the synthetic coordinate. To our knowledge, these measurements provide the first experimental evidence for the position-dependent localization structure predicted for the CST--AAH model and establish a spatially varying kinetic-energy scale as a programmable control axis for quasiperiodic localization. The platform opens a route to investigating interaction effects, dynamical scaling, and nonequilibrium transport in spatially inhomogeneous synthetic quantum systems.

This research is supported by the Guangdong-Hong Kong-Macau Greater Bay Area Quantum Science Center and funded through the Guangdong Province Quantum Science Strategic Initiative (Grant No.~GDZX2302005), Guangzhou-HKUST(GZ) Joint Funding Program (No.~2025A03J3783) and Guangzhou Municipal Science and Technology Project (No.~2025A04J7077, No.~2025A03J3861). 

\textit{Data availability}---The data that support the findings of this work are available from the corresponding author upon reasonable request.

\bibliography{apssamp}

\end{document}